\documentclass[a4paper,12pt]{article}
\usepackage{amsmath,amssymb,amsthm,mathrsfs,graphicx,float,colordvi,url,wrapfig,color}

\def\tf{\tilde{f}}  
  
\def\tm{\tilde{m}} 
\def\tF{\tilde{F}}

\def\tT{\tilde{T}}  
\def\tN{\tilde{N}} 
\def\tD{\tilde{\Delta}}

\begin{document}
\renewcommand{\thefootnote}{\fnsymbol{footnote}} 
\begin{titlepage}

\vspace*{5mm}

\begin{center}
{\large \bf {Hawking fluxes and Anomalies} 
\\ \vspace*{0.20 cm}
{in Rotating Regular Black Holes with a Time-Delay}
}
 
\vspace*{10mm}

\normalsize
{\large {Shingo Takeuchi\footnote{shingo(at)nu.ac.th}}} 

\vspace*{6mm}

\textit{
The Institute for Fundamental Study, ``The Tah Poe Academia Institute''},\\
\vspace*{0.10 cm}
\textit{Naresuan University Phitsanulok 65000, Thailand}\\

\end{center}

\vspace*{5mm}
 
\begin{abstract}
Based on the anomaly cancellation method we are going to compute the Hawking fluxes (the Hawking thermal flux and the total flux of energy-momentum tensor) 
from a four-dimensional rotating regular black hole with a time-delay. 
To this purpose, in the three metrics proposed in Ref.\cite{DeLorenzo:2015taa}, 
we try to perform the dimensional reduction in which the anomaly cancellation method is feasible at the near-horizon region in a general scalar field theory.  
As a result we can demonstrate that the dimensional reduction is possible in two of those metrics.  
Hence we perform the anomaly cancellation method and compute the Hawking fluxes in those two metrics. 
Our Hawking fluxes involve the three effects: 
1)~the quantum gravity effect regularizing the core of the black holes, 
2)~rotation of the black hole, 
3)~the time-delay. 
Further in this paper toward the metric in which the dimensional could not be performed, 
we argue that it would be some problematic metric, and mention its cause.  
The Hawking fluxes we compute in this study could be considered to correspond to more realistic Hawking fluxes. 
Further what Hawking fluxes can be obtained from the anomaly cancellation method would be interesting 
in terms of the relation between a consistency of quantum field theories and black hole thermodynamics. 
\end{abstract}
\end{titlepage}

\newpage 

\section{Introduction} 
Matters to be the gravity sources of the black holes form the singularities in the classical theories.  
However the quantum gravity effect working among those matters is considered to become stronger as those matters close up to the plank scale, 
and it is expected that the repulsive force preventing those matters from forming a singularity appears from the quantum gravity effect. 
As a result the singularities in the classical black holes can be expected not to present if the quantum gravity effect is taken into account.    
Those black holes are named {\it regular black hole} or {\it non-singular black hole}.  
Currently many stars expected as black holes have been observed, and the regular black holes are considered to be able to correspond to those actual black holes somehow.  
Hence regular black holes would be interesting both theoretically and phenomenologically.

Until now the Schwarzschild regular black hole and its rotating version have been proposed 
in Refs.\cite{Bardeen1968, Dymnikova:1992ux, AyonBeato:1998ub, Dymnikova:2001fb, Cho:2000kr, Hayward:2005gi, Nicolini:2005vd, Ansoldi:2006vg, DeLorenzo:2014pta}
and~\cite{Bambi:2013ufa, Toshmatov:2014nya, Ghosh:2014hea, Neves:2014aba} respectively.     
The four-dimensional rotating regular black holes with a time-delay~(a gravitational time dilation)~\cite{Shapiro:1964uw} has also been proposed~\cite{DeLorenzo:2015taa}.

The time-delay is an indispensable in the descriptions of any black holes. 
In classical black holes with singularities, the time-delay naturally appears. 
Hence we do not need to give attention particularly to the time-delay in classical black holes with singularities.  
However in the regular black holes, the time-delay becomes absent for a byproduct in the formalism for the regularization of the singularity, 
which we explain in appendix.\ref{app:TimeDelay} in this paper.     
Hence we need to give attention to the time-delay in the case of regular black holes.   
\newline

One of the most important problems in the black hole physics would be the information paradox and the microscopic unveiling of the Bekenstein-Hawking entropy obtained from the thermodynamic analogy.  
In those problems the Hawking fluxes play very important role.  
Hence to investigate the Hawking fluxes in the regular black holes would be very interesting.

So far several ways for the computation of the Hawking fluxes have been invented~(\cite{Umetsu:2010ts} as a review),  
and in 2005 a computation algorithm has newly been invented~\cite{Robinson:2005pd}.  
It is referred to as {\it anomaly cancellation method} or just {\it anomaly cancellation}, 
which is an algorithm to determine the Hawking fluxes based on a demand that there should be no anomalies in the quantum field theories at the near-horizon region.

The anomaly cancellation method could shed a new light to the microscopic unveiling of the black hole problems.  
Because in the anomaly cancellation method, 
a consistency in the microscopic description based on the quantum field theory~(the invariance under the local transformations at the quantum level) and Hawking fluxes are linked each other.

The algorithm of the anomaly cancellation method is feasible 
when the field theories can be reduced to an infinite collection of the two-dimensional free massless model at the near-horizon region.   
Once the field theories are reduced to such an infinite collection of the two-dimensional free model at the near-horizon region,  
the computation can be performed without any details of the original model. 
Since many field theories can be reduced to an infinite collection of two-dimensional free massless model at the near-horizon region in various kinds of black holes, 
the anomaly cancellation method can be considered as applicable in wide range of theories, 
and we can see that the Hawking fluxes are universal phenomena independent of the details of each theory. 
\newline

Hence we in this paper compute the Hawking thermal flux and the total flux of energy-momentum tensor\footnote{
For the terminologies, ``the Hawking thermal flux'' and ``the total flux of energy-momentum tensor'', we have followed Ref.\cite{Iso:2006ut}.} 
from the rotating regular black hole with a time-delay~\cite{DeLorenzo:2015taa} 
based on the anomaly cancellation method~\cite{Iso:2006wa,Iso:2006ut}.  
The computation in the case of the regular schwartzshild black holes has already been performed~\cite{Kim:2008hm}.  
However the computation involving the rotation and the time-delay has not been performed, and what we compute in this paper is that case.    
Further, if the black holes are regular black holes, the time-delay should be involved for the reason mentioned above.    
So in this study it would be considered that 
we compute a more realistic Hawking fluxes with the three effects:~1)~the quantum gravity effect regularizing the gravity source of the black holes, 
2)~the rotation of the black hole, and 3)~the time-delay.  
\newline

Organization of this paper is as follows:~ 
In Sec.\ref{Chap:RBHTD}, the metric of rotating regular black hole and three ways to involve the time-delay into it are given.  
In Sec.\ref{Chap:DRA}, 
taking a general scalar field theory,  
it is shown that the action of the matter fields can be reduced to an infinite collection of the two-dimensional free model near the horizon in the type II or III metrics. 
Further it is  argued that the type I metric in which the dimensional cannot be performed would be some problematic metric, and its cause is mentioned. 
In Sec.\ref{The regularization effect}, 
the Hawking fluxes (the Hawking thermal flux and the total flux of energy-momentum tensor) are computed 
based on the anomaly cancellation method.  
Sec.\ref{Summary} is the summary of this paper. 
In appendices, the regular black hole with a time-delay we consider in this paper is briefly reviewed, etc. 

\section{Rotating regular black holes with a time-delay}
\label{Chap:RBHTD} 
We consider the following rotating regular black hole metric\footnote{
The metric in eq.(\ref{RRBHTD}) is obtained~\cite{Bambi:2013ufa} from the Schwarzschild regular black hole~\cite{Hayward:2005gi} 
through the Newman-Janis algorithm~\cite{Newman:1965tw,Newman:1965my,Drake:1998gf}.}, 
\begin{eqnarray}\label{RRBHTD}
ds^2 = g_{tt} dt^2 + 2 g_{t\phi} dt d\phi + g_{\phi\phi}d\phi^2 + g_{rr}dr^2 + g_{\theta\theta}d\theta^2, 
\end{eqnarray}
with
\begin{eqnarray}\label{RR metric2}
g_{MN}=
\begin{pmatrix} 
-\frac{\tD - a^2 \sin^2\theta}{\Sigma}                     & 0                  & 0      & -\left(\frac{r^2+a^2 - \tD}{\Sigma} \right) a\sin^2 \theta \\  
0                                                          & \frac{\Sigma}{\tD} & 0      & 0 \\ 
0                                                          & 0                  & \Sigma & 0 \\  
-\left(\frac{r^2+a^2 - \tD}{\Sigma} \right) a\sin^2 \theta & 0                  & 0      &  \left( (r^2 + a^2)^2 - \tD a^2 \sin^2 \theta \right) \frac{\sin^2 \theta}{\Sigma}
\end{pmatrix},
\end{eqnarray} 
where $M,N = t, r, \theta, \phi$,  and
\begin{eqnarray}
\label{SigmatD}
\Sigma &=& r^2+a^2 \cos^2 \theta,\\
\label{tD}
\tD    &=& r^2 -2 \tm r +a^2,
\end{eqnarray} 
and
\begin{eqnarray}\label{MinRBH}
\tm = \frac{m_0 r^3}{r^3 + l_p^3}. 
\end{eqnarray}
$m_0$ is the ADM mass, $a = L/m_0$ where $L$ is the angular momentum of the black hole,  
and $l_p$ is the Plank length which originates in the quantum gravity effect regularizing the gravity source of the black hole.   
It tuns out that $\tD = 0$ is a fifth-order equation. So we cannot get $r_\pm$ analytically, 
where $r_\pm$ are the outer and inner horizon radii.  
Further, solving $r_\pm$ numerically is technically hard for the fact that $r_\pm$ depend on three parameters: $a$, $l_p$ and $m_0$. 
Hence in this study, expanding $\tm$ in terms of $l_p$ as $\tm = m_0 + \frac{m_0}{r^3} \, l_p^3 + \cdots$, 
and considering $l_p \ll 1$, we solve $\tD = 0$ up to its first-order.
In that case, $\tD$ can be written as $\tD = a^2 - 2m_0 r + r^2 + 2m_0 \, l_p^3/r^2 + O(l_p^6)$, where $O(l_p^6)$ is ignorable now, 
and $\tD = 0$ is given as a forth-order equation, which we formally write as
$0 = (r - r_{\rm I+})(r - r_{\rm I-})(r-r_+)(r -r_-)/r^2 + O(l_p^6)$. 
Finally the four solutions in $\tD = 0$ up to the first-order expansion of $l_p$ are 
\begin{eqnarray}
r_{\rm I\pm} &=& -\frac{2m_0^2}{a^4}l_p^3 + O(l_p^6),\\
\label{rpm}
r_\pm        &=& m_0 \pm \sqrt{m_0^2-a^2} + \frac{m_0}{a^4} \bigg( 2m_0 + \frac{\mp 2 m_0^2 \pm a^2}{\sqrt{m_0^2-a^2}} \bigg)l_p^3 + {\cal O}\big(l_p^6 \big). 
\end{eqnarray}
In $r_\pm$, let us note that the first two terms are the typical ones in the classical Kerr black holes, 
and the third term is the corrections coming from the quantum gravity effect.
So it is obvious that $r_{\rm I\pm}$ are unphysical solutions and $r_\pm$ are physical solutions. 

\begin{figure}[!h]  
\begin{center}   
\includegraphics[width=60.0mm,clip]{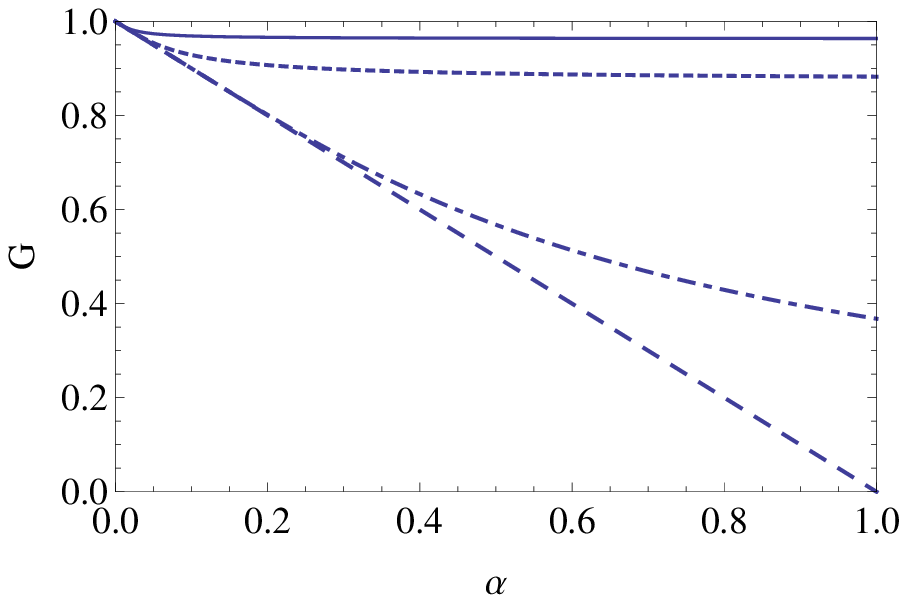} 
\includegraphics[width=60.0mm,clip]{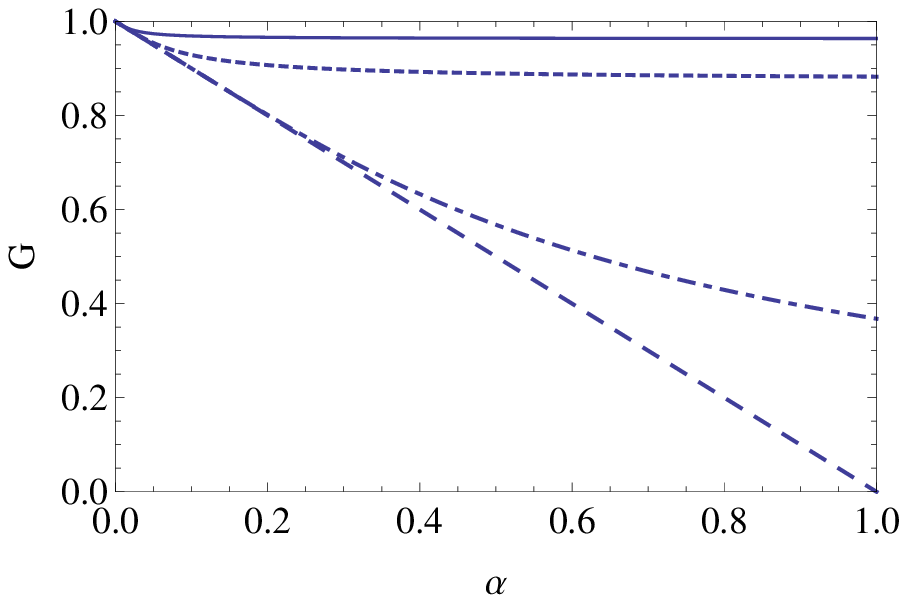} 
\end{center}
\caption{
{\it Left}\,:~The solid, dotted, chain and dashed lines are plotted with $\beta$ = 0.05, 0.50, 1.00 and 1.50, respectively, where $(m_0,r)=(1,1)$ constant. 
{\it Right}\,:~The solid, dotted, chain and dashed lines are plotted with $r$ = 3.00, 2.00, 1.00 and 0.05, respectively, where $(m_0,\beta)=(1,1)$ constant. 
}
\label{Fig_G}
\end{figure}

Then, we consider the three rotating regular black holes with a time-delay~\cite{DeLorenzo:2015taa}, 
which can be obtained by the following replacement in the metric of eq.(\ref{RRBHTD}): 
\begin{enumerate} 
\renewcommand{\theenumi}{(\Roman{enumi})} 
\item $g_{tt} \mapsto G \, g_{tt}$, modifying only the $dt^2$ term in the metric. 
\item $dt \mapsto \sqrt{G} \, dt$, modifying the $dt^2$ and $dtd\phi$ terms.
\item $dt \mapsto \sqrt{G} \, dt$ and $d\phi \mapsto \sqrt{G} \, d\phi$, modifying the $dt^2$, $dtd\phi$ and $d\phi^2$ terms.
\end{enumerate}
In this paper we refer to these three as the type~I, II and III metrics, and

\begin{eqnarray}\label{GTD}
G = 1-\alpha\left\{1-\exp \left(-\frac{\beta m_0}{\alpha r^3}\right) \right\}  
\end{eqnarray}
as the factor to involve a time-delay. 
$\alpha\, \in\, [0,~1)$ is the parameter to control the time-delay, and $\beta$ is some numerical constant associated with the 1-loop correction of the quantum gravity.  
$\alpha\sim 1$ corresponds to the situation that the time-delay is small. 
On the other hand $\alpha\sim 0$ corresponds to the situation that the time-delay is large~(For the explanation of this, see eq.(\ref{Ggttalpha})).  
We plot the $r$-, $\beta$- and $m_0$-dependences in the function $G$ in Fig.\ref{Fig_G}.  
Since the $\beta$- and the $m_0$-dependences are the same in the expression of $G$, we do not make a plot for the $m_0$-dependence in Fig.\ref{Fig_G}. We can see from Fig.\ref{Fig_G} that:
\begin{itemize}
\item  
$G$ becomes smaller as $\beta$ or $m_0$ become larger, for a same $\alpha$. 
\item
$G$ becomes smaller as $r$ becomes smaller, for a same $\alpha$. 
\end{itemize}

\section{Dimensional reduction of field theories near the horizon}
\label{Chap:DRA}
Now we consider the quantum field theories which will be the ingredient of the Hawking fluxes on the background space-times given as the type I, II and III metrics.  
For now, assuming that we have some field theory, let us focus on the scalar field part in the whole action, which we write as  
\begin{eqnarray}\label{scalar}
S_{\rm scalar} = \frac{1}{2} \, \int d^4x \, \sqrt{-g} \, \Big( g^{MN} \partial_M \phi^{\ast} \partial_N \phi + {\cal L}_{\rm mass} + {\cal L}_{\rm int} \Big).   
\end{eqnarray}

We can find that the term ${\cal L}_{\rm mass} + {\cal L}_{\rm int}$ is suppressed at the near-horizon region for the fact that $\tilde{\Delta}$ vanishes at the horizon.  
This suppression can be found by changing the radial coordinate to the tortoise coordinate $r^*$ defined as $d r^*/ d r = (r^2+a^2)/\tD$.  
Hence we disregard $\displaystyle {\cal L}_{\rm mass} + {\cal L}_{\rm int}$ at the near-horizon region.  
Then let us compute $\displaystyle \int d^4x \, \sqrt{-g} \, g^{MN}\partial_M \phi^{\ast} \partial_N \phi$ in each type of metric in what follows.
\newline

First, we compute $\sqrt{-g}$ in the type I, II and III metrics, which are 
\begin{eqnarray}\label{sqrtmg}
\sqrt{-g} = 
\left\{
\begin{array}{l}
\displaystyle a \left(a^2+r^2\right)  \sqrt{\frac{1-G}{\tD}} \sin ^2 \theta \quad \! \textrm{for the type I metric}, \vspace{4mm}\\
\displaystyle \sqrt{G} \, \Sigma \, \sin \theta \quad \! \textrm{for the type II metric}, \vspace{4mm}\\
\displaystyle G  \, \Sigma \, \sin \theta \quad \! \textrm{for the type III metric,} 
\end{array}
\right.
\end{eqnarray} 
where we have used $a^2 \cos \, (2 \theta) + a^2 + 2 r^2 = 2 \Sigma$ and taken the leading in $\tD \sim 0$.  
We will use these two in the following computation as necessary.


Next $g^{MN}$ in each type of metric are given as
\begin{eqnarray}
\label{InvMetI}
g^{MN}_{\rm I}&=& 
\left(
\begin{array}{cccc}
 \frac{1}{\sin^2\theta} \frac{\Sigma}{a^2(-1+G)} & 0 & 0 & \frac{1}{\sin^2 \theta} \frac{\Sigma}{a(a^2+r^2)(-1+G)} \\
 0 & \frac{\tD}{\Sigma} & 0 & 0 \\
 0 & 0 & \frac{1}{\Sigma} & 0 \\
\frac{1}{\sin^2\theta} \frac{\Sigma}{a(a^2+r^2)(-1+G)} & 0 & 0 & \frac{1}{\sin^2\theta} \frac{G \Sigma}{(a^2+r^2)^2(-1+G)} \\
\end{array}
\right),\\ \vspace{1.5mm}
\label{InvMetII}
g^{MN}_{\rm II}&=&
\left(
\begin{array}{cccc}
 -\frac{\left(a^2+r^2\right)^2}{G \tD \Sigma} & 0 & 0 & -\frac{a (a^2+r^2)}{\sqrt{G} \tD \Sigma} \\
 0 & \frac{\tD}{\Sigma} & 0 & 0 \\
 0 & 0 & \frac{1}{\Sigma} & 0 \\
 -\frac{a (a^2 + r^2 )}{\sqrt{G} \tD \Sigma} & 0 & 0 & -\frac{a^2}{\tD \Sigma} \\
\end{array}
\right),\\ \vspace{3mm}
\label{InvMetIII}
g^{MN}_{\rm III}&=&
\left(
\begin{array}{cccc}
 -\frac{\left(a^2+r^2\right)^2-a^2 \sin^2\theta  \tD}{G \tD \Sigma} & 0 & 0 & -\frac{a (a^2+r^2)}{G \tD \Sigma} \\
 0 & \frac{\tD}{\Sigma} & 0 & 0 \\
 0 & 0 & \frac{1}{\Sigma} & 0 \\
 -\frac{a (a^2 + r^2)}{G \tD \Sigma} & 0 & 0 & -\frac{a^2}{G \tD \Sigma} \\
\end{array}
\right).
\end{eqnarray} 


Then in common with the type I, II and III metrics, 
\begin{eqnarray}
\int d^4x \sqrt{-g} \, {\cal L} 
\!\! &=& \!\!
\frac{1}{2} \, \int d^4x 
\sqrt{-g} g^{MN} \partial_M \phi^\ast \partial_N \phi \nonumber \\
\!\! &=& \!\! 
- \frac{1}{2} \int d^4x \sqrt{-g}\,
\bigg[ \phi^\ast \Big\{ (\sqrt{-g})^{-1}\partial_\theta(\sqrt{-g}g^{\theta\theta}\partial_\theta)+g^{\phi\phi} \, \partial^2_\phi \Big\} \,\phi  \nonumber \\
&& \qquad \!\!\!
+ \, \phi^\ast \Big\{ g^{tt} \partial_t^2 + (\sqrt{-g})^{-1}\partial_r(\sqrt{-g}g^{rr}\partial_r)+2g^{\phi t} \, \partial_\phi \partial_t \Big\} \,\phi \, \bigg]. 
\end{eqnarray} 
Each part in the above can be computed as 
\begin{eqnarray}
\label{GIIIAngleP0}
&&
\int d^4x \sqrt{-g}\,
\phi^\ast 
\Big\{ (\sqrt{-g})^{-1}\partial_\theta(\sqrt{-g}g^{\theta\theta}\,\partial_\theta)+g^{\phi\phi}\,\partial^2_\phi \Big\} \, 
\phi  
\\ 
\label{GIIIAngleP}
&=&
\left\{
\begin{array}{l} 
\displaystyle
\int d^4x\, 
\phi^\ast \,
a(a^2+r^2) \sqrt{\frac{1-G}{\tD}} 
\left\{
\partial_\theta \left( \frac{\sin^2\theta}{\Sigma} \partial_\theta \right) + \frac{G\,\Sigma}{(a^2+r^2)^2(-1+G)} \partial_\phi^2 
\right\}
\phi
\vspace{2mm}\\  
\textrm{for type I metric}
\vspace{5mm}\\
\displaystyle
-\int d^4x\, 
\phi^\ast \, \frac{a^2\sqrt{G}}{\tD} \sin \theta \,\partial_\phi^2 \, 
\phi \quad \textrm{for type II metric} 
\vspace{5mm} \\
\displaystyle 
-\int d^4x\, 
\phi^\ast \, \frac{a^2}{\tD} \sin \theta \,\partial_\phi^2 \, 
\phi \quad \textrm{for type III metric}
\end{array}
\right.
\end{eqnarray}  
\begin{eqnarray}
&& 
\int d^4x\, \sqrt{-g}\, \phi^\ast \Big\{ 
g^{tt} \partial_t^2 + (\sqrt{-g})^{-1}\,\partial_r(\sqrt{-g}g^{rr}\partial_r)+2g^{\phi t}\partial_\phi\partial_t 
\Big\} \, \phi \nonumber\\
&=&
\int d^4x\, \sqrt{-g}\, \phi^\ast g^{tt} \Big\{  
\partial_t^2 + (g^{tt} \sqrt{-g})^{-1}\,\partial_r(\sqrt{-g}g^{rr}\partial_r)+2\frac{g^{\phi t}}{g^{tt}}\partial_\phi\partial_t 
\Big\} \, \phi  \nonumber\\
&=&
\label{GIIIKinP0}
\int d^4x\, \sqrt{-g}\,\phi^\ast g^{tt} \Big\{ 
( \partial_t + \frac{g^{\phi t}}{g^{tt}}\partial_\phi )^2 -(\frac{g^{\phi t}}{g^{tt}}\partial_\phi)^2+ (g^{tt} \sqrt{-g})^{-1}\,\partial_r(\sqrt{-g}g^{rr}\partial_r) 
\Big\} \, \phi \vspace{4.5mm} \\
\label{GIIIKinP}
&=&
\left\{
\begin{array}{l}
\displaystyle 
\int d^4x\, \sin^2 \phi^\ast \, \partial_r \left( a(a^2+r^2) \sqrt{\frac{\tD \, (1-G)}{\Sigma}} \, \partial_r \right) \phi 
\quad \textrm{for type I metric,}
\vspace{5mm}\\
\displaystyle 
-\int d^4x\, \phi^\ast \,  \frac{\sin \theta}{\sqrt{G}\tD} 
\left\{ 
(a^2+r^2)^2\left( \partial_t + \frac{a \, \sqrt{G}}{a^2+r^2}\,\partial_\phi \right)^2 
- a^2G \,\partial_\phi^2 
- \sqrt{G}\,\tD\partial_r (G\tD\,\partial_r)
\right\} \phi
\vspace{2mm}\\  
\textrm{for type II metric,}
\vspace{5mm}\\
 \displaystyle 
-\int d^4x\, \phi^\ast \,  \frac{\sin \theta}{\tD}
\left\{  
(a^2+r^2)^2 \left( \partial_t + \frac{a}{a^2+r^2}\,\partial_\phi \right)^2 
- a^2 \partial_\phi^2 
- \tD\,\partial_r (G\tD\,\partial_r)
\right\} \phi 
\vspace{2mm}\\   
\textrm{for type III metric.}
\\
\end{array}
\right.\nonumber\\
\end{eqnarray}

Summing up eqs(\ref{GIIIAngleP}) and (\ref{GIIIKinP}), for the type I metric,
\begin{eqnarray}\label{KI}
\int d^4x \sqrt{-g} \, {\cal L} \!\! &=& \!\!
- \frac{1}{2} \, \int d^4x \, \phi^\ast 
\left[
\sin^2\theta \, \partial_r\left( a(a^2+r^2) \sqrt{\frac{\tD(1-G)}{\Sigma}} \, \partial_r \right)
\right.
\nonumber\\ && 
\hspace{-15mm}
+a(a^2+r^2) \sqrt{\frac{1-G}{\tD}}
\left\{ 
\partial_\theta \left( \frac{\sin^2 \theta}{\Sigma}\partial_\theta \right) + \frac{G \, \Sigma}{(a^2+r^2)^2(1-G)} \partial_\phi^2 
\right\} 
\Bigg] \phi. 
\end{eqnarray}
On the other hand, in the type II and III metrics, 
\begin{eqnarray}\label{KIIaIII}
&&
\int d^4x \sqrt{-g} \, {\cal L}  
\nonumber \\
\!\!&=&\!\!
\frac{1}{2} \, \int d^4x  \, \phi^\ast
(a^2+r^2) \, G^s \sin \theta 
\left\{ 
\frac{a^2+r^2}{G^p \tD} \left( \partial_t +  \frac{a\,G^q}{a^2+r^2} \, \partial_\phi \right) 
- 
\partial_r \left( \frac{G^p\tD}{a^2+r^2} \partial_r \right)
\right\} \phi, \nonumber\\
\end{eqnarray}
where 
\begin{eqnarray}\label{pq}
\begin{array}{l}
(p,q,s) \,\,=\,\, (3/4,~1/2,~1/4) \quad \textrm{for the type II metric,} \vspace{2.5mm}\\
(p,q,s) \,\,=\,\, (1/2,~0,~1/2) \quad \textrm{for the type III metric.}
\end{array}
\end{eqnarray} 
In the above computations, for example to reach eq.(\ref{KIIaIII}) in the type II metric, we have performed the following computation:
\begin{eqnarray}
\!\! && \!\! 
\frac{\sin \theta}{\sqrt{G}\tD}  
\left\{ 
(a^2+r^2)^2 
\left( \partial_t + \frac{a\sqrt{G}}{a^2+r^2}\partial_\phi \right)^2 
- a^2\,G \partial_\phi^2 
- \sqrt{G}\,\tD\partial_r (G\tD\partial_r)
\right\} 
+
\frac{ a^2 \sqrt{G}}{\tD} \sin \theta \, \partial^2_\phi
\nonumber\\
\!\! &=& \!\! 
(a^2+r^2) \sin \theta 
\left\{ 
\frac{a^2+r^2}{\sqrt{G}\tD} 
\left( 
\partial_t + \frac{a\sqrt{G}}{a^2+r^2}\partial_\phi \right)^2 
-
\frac{1}{a^2+r^2}
\partial_r 
\left( G \tD \partial_r \right) 
\right\} 
\nonumber\\
\!\! &=& \!\! 
(a^2+r^2) \sin \theta \left\{ 
\frac{a^2+r^2}{\sqrt{G}\tD}  \left(  \partial_t + \frac{a\sqrt{G}}{a^2+r^2}\,\partial_\phi \right)^2 
- G^p \, \partial_r \left( \frac{G^{1-p} \tD}{a^2+r^2} \partial_r \right) 
\right\} 
\nonumber\\
\!\! &=& \!\! 
(a^2+r^2) \, G^p \sin \theta 
\left\{
\frac{a^2+r^2}{G^{1/2+p}\tD} \left(  \partial_t + \frac{a\sqrt{G}}{a^2+r^2}\,\partial_\phi \right)^2 
- \partial_r \left( \frac{G^{1-p} \tD}{a^2+r^2} \partial_r \right) 
\right\} 
\nonumber\\
\!\! &=& \!\! (a^2+r^2) \, G^{1/4} \sin \theta 
\left\{
\frac{a^2+r^2}{G^{3/4}\tD}   \left(  \partial_t + \frac{a\sqrt{G}}{a^2+r^2}\,\partial_\phi \right)^2 
- \partial_r \left( \frac{G^{3/4} \tD}{a^2+r^2} \partial_r \right) 
\right\},
\end{eqnarray}
where in the above description we have omitted to write the integral and $\phi^\ast$, $\phi$. 
From the second to third lines, we have performed the following computation:~First once rewriting the second term as
\begin{eqnarray}
\frac{1}{a^2+r^2}
\partial_r 
\left( G \tD \partial_r \right)
=
\partial_r \left( \frac{1}{a^2+r^2}  \right) G \tD \partial_r
+
\frac{1}{a^2+r^2} \partial_r \left( G \tD \partial_r \right)
=
\partial_r \left( \frac{G \tD}{a^2+r^2} \partial_r \right),  
\end{eqnarray}
and then
\begin{eqnarray}
    \partial_r \left( \frac{G \tD}{a^2+r^2} \partial_r \right)
  \! &=& \! \partial_r \left( \frac{G^p G^{1-p} \tD}{a^2+r^2} \partial_r \right) 
  \,\,=\,\,   \partial_r (G^p) \, \frac{G^{1-p} \tD}{a^2+r^2} \partial_r + G^p \partial_r \left( \frac{G^{1-p} \tD}{a^2+r^2} \partial_r \right) \nonumber\\
  \! &=& \! G^p \partial_r \left( \frac{G^{1-p} \tD}{a^2+r^2} \partial_r \right).  
\end{eqnarray}
From the forth to fifth lines, from the demand $1/2+p=1-p$, we have fixed $p$ to $1/4$.
\newline

Continuing the computation in the type II or III metrics, 
expanding the scalar field by the spherical harmonic oscillator function:~$\phi(t,r,\theta,\phi)=\sum_{lm}\varphi_{lm}(t,r)Y_{lm}(\theta, \phi)$, 
and using the eigen equation for the angular momentum operator and the inner product: 
\begin{eqnarray}\label{formulaeY}
-i \, \partial_\phi Y_{lm}=L_z \, Y_{lm}=m \, Y_{lm}  
\quad {\rm and} \quad 
\int d\theta d\phi \, \sin \theta \, Y^\ast_{l'm'}Y_{lm}=\delta_{l'l}\,\delta_{m'm},
\end{eqnarray}
we can find that the models reduce to infinite free massless scalar field theories on the two-dimensional space of $(t,r)$ as
\begin{eqnarray}\label{2DRA}
S_{\rm scalar} 
= - \frac{1}{2}\,\sum_{lm} \int dt dr \, \Phi \, \varphi_{lm}^* \Big( g^{tt} \left( \partial_t + im A_t \right)^2 + \partial_r g^{rr} \partial_r \Big) \varphi_{lm},
\end{eqnarray}
with the effective U(1) gauge field and dilaton as
\begin{eqnarray}
\label{MetricsType}
(g_{tt},g_{rr}) &=& \bigg(-G^p \tf,\,\frac{1}{G^p \tf}\bigg),\quad \tf \equiv \frac{\tD}{a^2+r^2} \\
   (A_t,\, A_r) &=& \Big(-\frac{a\,G^q}{a^2+r^2},\, 0 \Big),\nonumber\\
           \Phi &=& (r^2+a^2)\,G^s,\nonumber 
\end{eqnarray}
where $p$, $q$ and $s$ are given in eq.(\ref{pq}). 

On the other hand, it can be seen that 
the scalar field in the type I metric cannot reduce to the two-dimensional theory at the near-horizon region as can be seen from eq.(\ref{KI}).  
\newline

Here let us consider what the dimensional reduction occurs at the near-horizon region intuitively. 

The motions of particles\footnote{
In this discussion we take the particle-picture's point of view, 
but it is the same if we discuss in terms of the propagation of the fields. 
Here in Ref.\cite{Parikh:1999mf} 
there is a description that taking the particle-picture  at the near horizon region would be valid, 
and actually based on the particle-picture, the analysis in Ref.\cite{Parikh:1999mf} goes ahead using the WKB approximation.
} 
which the field theory represents at the near-horizon region are composed of the longitudinal and transverse directions toward the horizon.  
Then since the strong attractive gravity force toward the horizon works in the longitudinal direction, 
the motions in the longitudinal direction would become much larger than the motions in the transverse direction. 
As a result the motions in the transverse direction become negligible at the near-horizon region. 
Hence it would be considered that what the dimensional reduction occurs at the near-horizon region would be natural.  
(At this time the motions in the longitudinal direction become the ingoing modes only.)   

Next at the near-horizon region, 
the motions of particles become mostly a free falling with the speed of light. 
As a result the kinetic energy in the particles becomes maximally large. 
Hence in the motions with a finite mass at the near-horizon region, 
what the mass term becomes negligible would be natural.  
Actually the mass term in the model we consider in this study at the near-horizon region is finite, as in eq.(\ref{scalar})\footnote{
Here this finite mass is the one after becoming larger by getting the relativistic effect.  
Hence this mass is considered as very small in the place except for the near-horizon region. 
Hence we could say that in our anomaly cancellation method, 
we are considering only the very light modes which get enlarged no more than some finite values even at the near-horizon region as in eq.(\ref{scalar}), 
and modes with mass which becomes comparable with the kinetic term at the near-horizon region are not considered.
}
.

Lastly, since all the modes uniquely perform mostly a free falling with speed of light as the massless modes at the near-horizon region,   
it would be considered that the interaction effects among them vanish. 
Hence what the interaction term vanishes would be natural.    
\newline   

So from the above, it would be considered that what the dimensional reduction occurs is natural, 
and the type I metric in which the dimensional reduction cannot be performed is considered as  some problematic metric.

The reason for the feasibility in the dimensional reduction in type II and III metrics and the infeasibility in type I metric is as follows: 
\begin{itemize}
\item 
In order for the dimensional reduction to occur, 
we can see that one of the necessary conditions is that 
the contribution of the angular part in eq.(\ref{GIIIAngleP0}) and the term concerning the angle, $-(g^{\phi t}/g^{tt} \partial_\phi)^2$, in eq.(\ref{GIIIKinP0}) should be canceled each other. 

This canceling can be possible in the case of the type II and III metrics. 
Because  
$g^{\theta\theta}\propto 1$ and $g^{\phi\phi}\propto \tD^{-1}$ as in eqs.(\ref{InvMetII}) and (\ref{InvMetIII}). 
As a result, in eq.(\ref{GIIIAngleP0}) the second term of $g^{\phi\phi}$ becomes dominant at the near-horizon region where $\tD\sim0$, and can survive, 
while the first term negligible. 
Then since $g^{\phi\phi}\partial_\phi^2 = g^{tt}(g^{\phi t}/g^{tt} \partial_\phi)^2$ can be valid in the case of the type II and III metrics,  
where these $g^{\phi\phi}\partial_\phi^2$ and $g^{tt}(g^{\phi t}/g^{tt} \partial_\phi)^2$ mean the coefficients in eqs.(\ref{GIIIAngleP0}) and (\ref{GIIIKinP0}) respectively, 
the contribution of the angular part in eq.(\ref{GIIIAngleP0}) and the term concerning $-g^{tt}(g^{\phi t}/g^{tt} \partial_\phi)^2$ in eq.(\ref{GIIIKinP0}) can be canceled each other in the case of the type II and III metrics.  
 
On the other hand, in the case of the type I metric,   
$g^{\theta\theta} \propto 1$ and $g^{\phi\phi} \propto 1$ in terms of the order of $\tD$ as in eq.(\ref{InvMetI}). 
As a result, the two terms in eq.(\ref{GIIIAngleP0}) become comparative each other at the near-horizon region, and can survive at the near-horizon region.   
However in the case of the type I metric, these two terms cannot be canceled by the term concerning $-g^{tt}(g^{\phi t}/g^{tt} \partial_\phi)^2$ in eq.(\ref{GIIIKinP0}).

\item 
Another one reason which makes the dimensional reduction feasible in the type II and III metrics is that 
the invariant volume $\sqrt{-g}$ is proportional to $\sin \theta$ in the case of the type II and III metrics as in eq.(\ref{sqrtmg}). 
As a result the integral meaning the inner product of $Y_{lm}$ in eq.(\ref{formulaeY}) can become feasible in the computation in the case of the type II and III metrics.

In contrast, from this we can see that the dimensional reduction is infeasible in the case of the type I metric, 
since the invariant volume is proportional to $\sin^2 \theta$  as in eq.(\ref{sqrtmg}) in the case of the type I metric. 
\end{itemize}

From this result, we can suppose that 
if the background space-time is the type II or III metric, 
the whole field theory action considered in the beginning of this section can also get the dimensional reduction 
to an infinite two-dimensional free massless action like eq.(\ref{2DRA}) near the horizon. 
Since the dimensional reduction like eq.(\ref{2DRA}) is indispensable in the anomaly cancellation method, 
we in this paper consider the type II and III metrics in the computation for the Hawking fluxes performed in what follows.

Here let us note that we do not need to specify the model of the field theory~(e.g. how mass and interaction terms are etc).  
Because we in this paper consider the situation after the models have reduced to the two dimensional theory.  
even if whatever the models before the dimensional reduction are. 
\newline

Now that the expressions near the horizon have been obtained as in eq.(\ref{2DRA}), 
let us here give the surface gravities in the type II and III metrics as
\begin{eqnarray}\label{surfaceG}
\kappa = \frac{2\pi}{\beta} = \frac{1}{2}\partial_r \big( G^p\tf \big) \big|_{r=r_+} = \frac{1}{2}\partial_r \tF_p \big|_{r=r_+},~ {\rm where} \quad \tF_p \equiv G^p\tf.
\end{eqnarray} 
We can see that this is consistent with the one in Ref.\cite{Kim:2008hm} when $a=0$~(no rotating case) and $G=1$~(no time-delay case).

Further for the convenience in the actual computation, 
we also give the components of the Christoffel in the background in eq.(\ref{2DRA}) as 
\begin{eqnarray}\label{Christoffel}
\Gamma^t_{\mu\nu}=
\begin{pmatrix}
0 && \tF_p'/(2\tF_p) \\
\tF_p'/(2\tF_p) && 0 
\end{pmatrix}, \quad
\Gamma^r_{\mu\nu}=
\begin{pmatrix}
\tF_p \tF_p'/2 && 0 \\
0 && -\tF_p'/(2\tF_p) 
\end{pmatrix},
\end{eqnarray}
where $\mu,\nu=t,r$.

\section{Computation of the Hawking fluxes from the anomaly cancellation}
\label{The regularization effect}
We in this section compute the U(1) gauge current and the energy-momentum tensor in general quantum field theories 
having reduced to infinite two-dimensional free massless models at the near horizon region in the type II and III metrics 
based on the anomaly cancellation method\footnote{
A description on the relation between the U(1) gauge currents and the energy-momentum tensors 
in the two-dimensional and four-dimensional field theories in classical Kerr black holes with singularities is given in Ref.\cite{Iso:2006ut}.}.  
Then we finally obtain the Hawking thermal flux and the total flux of energy-momentum tensor from the rotating regular black hole with a time-delay.

Our computation is carried out in the same fashion with Refs.\cite{Iso:2006wa,Iso:2006ut} basically. 
However in our analysis, $\epsilon$ and $r_0$ appear in eqs.(\ref{region_H}) and (\ref{region_O}). 
Those parameters do not appear in Refs.\cite{Iso:2006wa,Iso:2006ut} and any other related papers. 
In using $\epsilon$ and $r_0$, the meaning of computation does not change. 
We use those just to give attention to the applicable limit for what both ingoing and outgoing modes present  or only ingoing modes present at the near-horizon region
and what the theories can be regarded as an infinite two-dimensional free massless model or not.     
For more detail, see the explanation under eqs.(\ref{region_H}) and (\ref{region_O}).

What follows are common to the type II and III metrics. 
So in what follows, we do not specify which computation and result correspond to which type of metrics.

\subsection{Dividing of the radial direction} 
\label{subsec:DRD}
First we set up the radial direction. 
It is known that on the horizon the outgoing modes do not present and there is the ingoing modes only. 
Then as a setting, corresponding to such a situation we divide the radial direction in the exterior of the horizon into two regions using a parameter $\epsilon$ as 
\begin{eqnarray}
\label{region_H}
&\bullet& \qquad \quad\!\! r_+         \leq  \, r \, \leq r_+ + \epsilon, \quad \textcolor{black}{\textrm{($\cal H$~:~there is only ingoing modes)}}  \\
\label{region_O}
&\bullet& \quad r_+ + \epsilon <  \, r \,  \leq r_o, \qquad \,\,\,\,\, \textcolor{black}{\textrm{($\cal O$~:~there are out and ingoing modes).}}
\end{eqnarray} 
Corresponding to this, $\epsilon$ is very small, $\epsilon \ll 1$. 
$r_{o}$ stands for the maximum $r$ in which the quantum field theories can be uniquely assumed to be the two-dimensional model. 
Namely we assume that the dimensional reduction in Sec.\ref{Chap:DRA} is possible for $r_+ \leq r \leq r_o$. 
We refer to the two regions in eqs.(\ref{region_H}) and (\ref{region_O}) as the regions $\cal H$ and $\cal O$, respectively. 
Here, the left- and right-handed modes in the two-dimensional field theories correspond to the outgoing and ingoing modes respectively.

Although now we have divided the radial direction using $\epsilon$ sharply, 
the conserved currents are also represented sharply using the step functions as in eqs.(\ref{tcu}) and (\ref{tmu}). 
This representation is a typical one in the analysis of the anomaly cancellation method~\cite{Iso:2006wa,Iso:2006ut}. 
Of course the representation using the step function is not the one naturally derived, but an artificial proposal. 
It should be a smooth function in nature. 
There are some attempts not using the step function~\cite{Banerjee:2007uc}~(This problem is also touched on in Chapter.4 in Ref.\cite{Umetsu:2010ts}). 
However in that case other unnatural stuffs arise.

\subsection{Hawking thermal flux from the computation of the U(1) gauge current}
\label{subsec:The angular momentum}
Corresponding to the regions $\cal O$ and $\cal H$, we write the U(1) gauge current as
\begin{eqnarray}\label{tcu}
J^\mu = J^\mu_{(o)} \Theta_+ + J^\mu_{(H)} H, 
\end{eqnarray} 
where $\mu,\nu=t,r$, and $\Theta_+ = \Theta(r-(r_+ + \epsilon))$ and $H = 1-\Theta_+$.  
Here $\Theta(r)$ is a step function.  
Thus $J^\mu_{(o)}$ and $J^\mu_{(H)}$ stand for the U(1) gauge current in the regions $\cal O$ and $\cal H$, respectively. 
Since our background space-time is independent of the time as in eq.(\ref{2DRA}),  
we let the U(1) gauge current have the $r$-dependence only:~$J^\mu_{(o)} = J^\mu_{(o)}(r)$ and $J^\mu_{(H)} = J^\mu_{(H)}(r)$.

In the region $\cal O$, the U(1) gauge current is conserved as 
\begin{eqnarray}\label{conservation low at O}
\nabla_\mu J^\mu_{(o)} = 0.
\end{eqnarray} 
On the other hand, in the region $\cal H$, 
let us consider the contribution of the left- or right-handed modes distinctively in the conservation equation of the U(1) gauge current. 
Then as the anomalous equation for the chiral transformation in the two-dimensional theories, it can be written as
\begin{eqnarray}\label{conservation low at H}
\nabla_\mu J^\mu_{(H)} = \pm \frac{m^2}{4\pi\sqrt{-g}} \epsilon^{\mu\nu}\partial_\mu A_\nu, 
\end{eqnarray}  
where $\epsilon^{tr}=1$, and $+/-$ respectively correspond to the chiral situation where there is only left-/right-handed modes, respectively.  
Since only the left-handed modes present in the region $\cal H$, we consider the case of the plus sign in eq.(\ref{conservation low at H}). 
Further we can find that $\nabla_\mu J^\mu_{(H)}=\nabla_t J^t_{(H)} + \nabla_r J^r_{(H)} = \partial_r J^r_{(H)}$ in the background in eq.(\ref{2DRA}). 
So l.h.s. of eq.(\ref{conservation low at H}) becomes $\displaystyle \partial_r J^r_{(H)}$. 
In addition from this, since it ends up in this study that there is no equation to determine the value of $J^t_{(H)}$, it is possible in this study to assume $J^t_{(H)}=0$.

Here if we compute the two-dimensional chiral anomaly~\cite{Bertlmann:Anomalies,Fujikawa:2004}, 
we can find that eq.(\ref{conservation low at H}) is given by $\nabla_\mu J^\mu_{(H)} = \pm \frac{m}{4\pi\sqrt{-g}} \epsilon^{\mu\nu}\partial_\mu A_\nu$.   
(This is because, briefly saying, in the case of $D=2$, in the expansion at eq.(4.60) in Ref.\cite{Fujikawa:2004} the first-order term survives instead of the second-order term. 
As a result the power of $e$ in eq.(4.61) in Ref.\cite{Fujikawa:2004} changes to $1$, where this $e$ corresponds to the $m$ in this study.) 
Hence there is apparently a discrepancy in the power of $m$. 
However our $U(1)$ gauge current $J^\mu_{(H)}$ could be considered to be defined as $J^\mu_{(H)} \equiv m \bar{\psi}\gamma^\mu \psi$.  
Here the usual one is $\bar{\psi}\gamma^\mu \psi$. 
The former and latter ones are interpreted as the electric charge density current and the probability density current, respectively.

Backing to eq.(\ref{conservation low at H}), hence we can write the U(1) gauge currents in the regions ${\cal O}$ and ${\cal H}$ as 
\begin{eqnarray}
\label{Ttr's1} 
J^r_{(o)} &=& c_o, \\
\label{Ttr's2} 
J^r_{(H)} &=& c_H + \frac{m^2}{4\pi} \int_{r_+}^r dr \partial_r A_t,
\end{eqnarray} 
where $c_o$ and $c_H$ are the integral constants that respectively stand for the values of the U(1) gauge currents at $r=r_o$ and $r_+$.  
$A_t$ in the type II and III metrics are given in eq.(\ref{MetricsType}). 
The $r$ in the upper bound of the integration is taken less than $r_o$. 
Here we do not need to consider the $t$-component of the current $J^t$.  
Because we can show $\nabla_\mu J^\mu=\partial_r J^r$. 
As a result, $J^t$ is irrelevant in the computation of eq.(\ref{dWU}).

Now let us consider the effective action $W$ given by the two-dimensional model at the near horizon region, 
in which all the modes in the interior of the horizon and the outside of $r_o$ have been integrated out.
Then, the variation of $W$ under the U(1) gauge transformation can be written as
\begin{eqnarray}\label{dWU}
-\delta W &=& \int dtdr \sqrt{-g} \,\lambda\, \nabla_\mu J^\mu \nonumber \\
          &=& \int dtdr \lambda  \left\{ \delta\big(r - (r_+ + \epsilon) \big) \left(J_{(o)}^r-J_{(H)}^r+\frac{m^2}{4\pi}A_t\right) 
                                                + \frac{e^2}{4\pi} \partial_r \left( A_t H \right) \right\},
\end{eqnarray} 
where $\lambda=\lambda(t,r)$ is the U(1) gauge transformation parameter, the integration range of $r$ is $r_+ \le r \le r_o$, 
and $J^\mu$ is given in eqs.(\ref{Ttr's1}) and (\ref{Ttr's2}).

The above computation leads the anomalous situation if r.h.s. in eq.(\ref{dWU}) does not vanish. 
The theory should be anomaly-free even near the horizon, and $\delta W$ must be zero.  
Here because of the quantum tunneling process, 
indeed there is the contribution of the the right-handed modes even near the horizon at the quantum level, 
which cancels the second term in r.h.s.\footnote{
The computation of the Hawking radiation based on the quantum tunneling processes was performed in Ref.\cite{Parikh:1999mf}.}.  
As a result the first term remains in r.h.s.. 
The first term should also vanish for the anomaly-free, which can be realized by the following constraint:
\begin{eqnarray}\label{ACCJ}
c_0 = c_H - \frac{m^2}{4\pi}A_t (r_+). 
\end{eqnarray}
By this we can consider that now $\delta W$ becomes zero and the situation becomes anomaly-free.

We now fix $c_H$ by the boundary condition that the covariant current~\cite{Bardeen:1984pm}: 
\begin{eqnarray}\label{Covariant current}
\tilde{J}^\mu = J^\mu \mp \frac{m^2}{4\pi\sqrt{-g}} \epsilon^{\nu\mu} A_\nu, 
\end{eqnarray}
vanishes at the horizon:~$\tilde{J}^\mu\big|_{r=r_+}=0$, where $-$ and $+$ respectively correspond to the left- and right-handed modes, 
and whose conservation equation is $\nabla_\mu \tilde{J}^\mu = \pm \frac{m^2}{4\pi\sqrt{-g}} \epsilon_{\mu\nu}F^{\mu\nu}$.  
As a result $c_H$ can be fixed as $c_H = - \frac{m^2}{4\pi}A_t (r_+)$.

Here there might be a question that, not considering $\tilde{J}^\mu$ only when we impose the boundary condition, 
but why all the computation had not performed using the covariant current from the beginning.  
Indeed if one considers $\tilde{J}^\mu$, since the conserved equation can be written in the covariant form, 
if we had performed our analysis with $\tilde{J}^\mu$ in eq.(\ref{Covariant current}) (and eq.(\ref{CovaAnomaly1})), 
our computation would have been covariant.
Of course what the covariant current is used only when the boundary condition is considered is a problem.  
But basically this is a typical procedure in the anomaly cancellation method~\cite{Iso:2006wa,Iso:2006ut}~(This problem is also touched on in Chapter.4 in Ref.\cite{Umetsu:2010ts}) 
as well as a problem in the representation using step functions in eq.(\ref{tcu}) (and (\ref{tmu}))~(also see the last paragraph in Sec.\ref{subsec:DRD}).  
Actually there is a study performing only with the covariant current~\cite{Banerjee:2007qs}. 
However in that case there arises other unnatural stuffs.

Then substituting the $c_H$ obtained above into $c_0$ in eq.(\ref{ACCJ}), we can obtain $c_0$ as   
\begin{eqnarray}\label{c0_AM_result}
c_0 = -\frac{m^2}{2\pi}A_t(r_+)=\frac{m^2a G^q(r_+)}{2\pi(r_+^2+a^2)}, 
\end{eqnarray}
where $q$ for the type II and III metrics are given in eq.(\ref{pq}). 
Since $q=0$ in the type III metric, $c_0$ in the type III metric has no dependence on the time-delay. 
Let us notice that $m$ appearing in the above expression is not mass but an index in the partial wave expansion in the above of eq.(\ref{2DRA}). 
\newline

From the same discussion in the Appendix.A in Ref.\cite{Iso:2006ut}, 
we can see that the above $c_0$ can be identified with the Hawking thermal flux from the regular rotating black hole with a time-delay. 
To show this identification we repeat the virtually same description with Ref.\cite{Iso:2006ut} in what follows.

In general the thermal fluxes from the rotating black holes can be computed as
\begin{eqnarray}\label{28}
m\int^\infty_0\frac{d\omega}{2\pi}\big( N_m(\omega) - N_{-m}(\omega)\big) =\frac{m^2 \Omega}{2\pi},
\end{eqnarray}
where $N_m(\omega)$ is the distribution function of fermions in the rotating black holes with a time-delay:
\begin{eqnarray}\label{DistFuncFer}
N_m(\omega) = \frac{1}{e^{\beta(\omega-m \Omega)}+1}  \quad {\rm with} \quad \Omega = \frac{a G^q(r_+)}{r_+^2+a^2}.
\end{eqnarray}
We can see that the result in eq.(\ref{28}) is consistent with our result in eq.(\ref{c0_AM_result}).

Our result involves the three effects:
\begin{itemize}
\item $G$~(The time-delay)
\item $a$~(The angular momentum of the black hole) 
\item $\l_p$~(The quantum gravity effect regularizing our black hole singularity via $r_+(\l_p)$ as in eq.(\ref{rpm}))
\end{itemize} 

We can see that when there is no time-delay~($G=1$) and no quantum gravity effect~($\l_p=0$), 
our result in eq.(\ref{c0_AM_result}) can agree with the Hawking thermal flux from the classical Kerr black hole given in eq.(25) in Ref.\cite{Iso:2006ut}.

The $r$-, $\beta$- and $m_0$-dependences in $c_0$ via $G$ can be known from Fig.\ref{Fig_G}. We can see from Fig.\ref{Fig_G} that: 
\begin{itemize}
\item  
When $\alpha$ much closes to zero, the differences in $r$-, $\beta$- and $m_0$-dependences in $c_0$ are little and $G \sim 1$. 
\item
When $\alpha$ is not near zero,
\begin{itemize}
\item $c_0$ becomes smaller as $\beta$ or $m_0$ gets larger with a constant $r$.
\item $c_0$ becomes smaller as $r$ gets smaller with a constant $\beta$ and $m_0$. 
\end{itemize}
\end{itemize}
\noindent
Here $\alpha\sim 0$ corresponds to the situation that the time-delay is large. 
On the other hand $\alpha\sim 1$ corresponds to the situation that the time-delay is small. 
For the relation between $\alpha$ and largeness of the time-delay, see eq.(\ref{Ggttalpha}).

\subsection{Total flux of energy-momentum tensor}
\label{subsec:The Hawking radiation}
Next we determine the energy-momentum tensor. 
In order to this, we start with the general coordinate transformation;~$x^\mu \mapsto x'{}^\mu = x^\mu - \xi^\mu(t,r)$. 
Corresponding to this, the fields are shifted, which are given by the Lie derivative:~$\delta g^{\mu\nu}=-(\nabla^\mu \xi^\nu+\nabla^\nu \xi^\mu)$, 
$\delta A_\mu = \xi^\nu \partial_\nu A_\mu +\partial_\mu \xi^\nu A_\nu$ and $\delta \Phi = \xi^\mu \partial_\mu \Phi$.       

Since $g_{\mu\nu}$, $A_\mu$ and $\Phi$ in our background are fixed as in eq.(\ref{2DRA}), 
the distribution function can be written as 
$Z = Z[g_{\mu\nu},A_\mu,\Phi]=\int{\cal D}[{\rm{matters}}] \exp\,{iS[g_{\mu\nu},A_\mu,\Phi,{\rm{matters}}]}$ in the region $r_+ \le r \le r_o$,  
without including $g_{\mu\nu}$, $A_\mu$ and $\Phi$ in the  path-integral.   
Here the scalar part in ${\cal D}[{\rm{matters}}]$ is ${\cal D}\varphi_{lm}$, where $\varphi_{lm}$ is given above eq.(\ref{2DRA}).   

Now let us consider the variation of $Z$ under the above general coordinate transformation.
Then since $Z$ should be invariant under the general coordinate transformation, we can write the following Ward identity: 
\begin{eqnarray}\label{coordid}
\left[ \delta g^{\mu\nu}\frac{\delta}{\delta g^{\mu\nu}} + \delta A_\mu \frac{\delta}{\delta A_\mu} +\delta \Phi \frac{\delta}{\delta \Phi}  \right] Z = 0.  
\end{eqnarray} 
Using the energy-momentum tensor $T_{\mu\nu} = \frac{2}{\sqrt{-g}}\frac{\delta S}{\delta g^{\mu\nu}}$ and current $J^\mu = \frac{1}{\sqrt{-g}}\frac{\delta S}{\delta A_\mu}$, 
from eq.(\ref{coordid}) we can obtain the following identity:  
\begin{eqnarray}\label{Anomaly2}
\nabla_\mu T^\mu{}_\nu 
&=& 
F_{\mu\nu}J^\mu  
+ A_\nu \nabla_\mu J^\mu 
-\frac{\partial_\nu \Phi}{\sqrt{-g}}\frac{\delta S}{\delta \Phi}
\pm \frac{1}{96\pi \sqrt{-g}}\epsilon^{\beta\delta} \partial_\delta \partial_\alpha \Gamma^\alpha_{\nu \beta}. 
\end{eqnarray} 
In the above, the last term is the consistent anomaly, 
and $+/-$ is adopted 
when there is only the left~(outgoing)/right~(ingoing) -handed modes, respectively. 
The second term is also an anomalous term when the situation is chiral, and its value is given in eq.(\ref{conservation low at H}). 
The first and third terms arise due to the fact that our background gauge and dilaton fields are classical as in eq.(\ref{MetricsType}). 
Here we ignore the term $-\frac{\partial_\nu \Phi}{\sqrt{-g}}\frac{\delta S}{\delta \Phi}$ in what follows. 
We state on this below.

In Ref.\cite{Iso:2006wa} and other related papers, 
the term $-\frac{\partial_\nu \Phi}{\sqrt{-g}}\frac{\delta S}{\delta \Phi}$ appears when the Ward identity is computed as in eq.(\ref{Anomaly2}). 
But the actual computation in the anomaly cancellation method is performed without the term $-\frac{\partial_\nu \Phi}{\sqrt{-g}}\frac{\delta S}{\delta \Phi}$. 
Its reason is stated in Ref.\cite{Iso:2006wa} as ``Since we are considering a static background, the contribution from the dilaton background can be dropped''.  
But it is difficult for me to understand the validity in ignoring the term $-\frac{\partial_\nu \Phi}{\sqrt{-g}}\frac{\delta S}{\delta \Phi}$ from this statement.   
In other related papers the computation is performed in the same fashion but its reason is also not stated clearly as far as I checked. 
However it seems that the dilaton $\Phi$ can be absorbed into $\varphi_{lm}$ by rescaling $\varphi_{lm}$ in eq.(\ref{2DRA}). 
Then since $\varphi_{lm}$ is the fields to get the path-integral, we could forget dilaton $\Phi$. 
Therefore we could ignore the term $-\frac{\partial_\nu \Phi}{\sqrt{-g}}\frac{\delta S}{\delta \Phi}$ in our computation.

Now as well as eq.(\ref{tcu}), we write the energy-momentum tensor as
\begin{eqnarray}\label{tmu}
T^\mu{}_\nu = T^\mu{}_{\nu(o)} \Theta_+ + T^\mu{}_{\nu(H)} H.  
\end{eqnarray} 
Then, in the region $\cal O$, the identity of eq.(\ref{Anomaly2}) leads to
\begin{eqnarray}\label{Grav Anomaly eq in O}
\nabla_\mu T^\mu{}_{\nu(o)} = F_{\mu\nu}J^\mu_{(o)}. 
\end{eqnarray} 
Hence, 
\begin{eqnarray}
T^r{}_{t(o)} = a_0+c_0 \int_{r_o}^r dr \partial_r A_t,
\end{eqnarray}
where we have used eq.(\ref{Ttr's1}). 
$a_0$ stands for the value of $T^r{}_{t}$ at $r=r_o$.  
Here, for the reason written below eq.(\ref{Grav Anomaly eq in H1}), we do not consider the components in $T^\mu{}_{\nu(o)}$ other than $T^r{}_{t(o)}$.

On the other hand, in the region $\cal H$, as well as the previous section, 
only left-handed modes present.    
Then the identity of eq.(\ref{Anomaly2}) leads to
\begin{eqnarray}\label{Grav Anomaly eq in H1}
\nabla_\mu T^\mu{}_{\nu(H)} = 
F_{\mu\nu}J^\mu_{(H)} 
+ A_\nu \nabla_\mu J^\mu_{(H)} 
+ \frac{1}{96\pi \sqrt{-g}}\epsilon^{\beta\delta} \partial_\delta \partial_\alpha \Gamma^\alpha_{\nu \beta}. 
\end{eqnarray} 
We can find that $\nabla_\mu T^\mu{}_{r(H)}=0$, 
where one of stuffs having been used in obtaining this is $J^t_{\rm (H)}=0$ which is mentioned under eq.(\ref{conservation low at H}). 
Hence let us turn to the case that $\nu=t$. 
In this case, we can find that $\nabla_\mu T^\mu{}_{t(H)}=\partial_r T^r{}_{t(H)}$.   
So from eq.(\ref{Grav Anomaly eq in H1}), 
\begin{eqnarray}\label{Grav Anomaly eq in H2}
\partial_r T^r{}_{t(H)} \!\! &=& \!\! F_{rt}J^r_{(H)} + A_t \partial_r J^r_{(H)} + \partial_r N^r{}_t\nonumber\\
                        \!\! &=& \!\! \partial_r \left( \frac{m^2}{4\pi}A_t^2 +c_0 A_t + N^r{}_t \right),
\end{eqnarray} 
where 
\begin{eqnarray}
N^r{}_t = \frac{1}{192\pi}(\tF_p'^2+\tF_p \tF_p''), 
\end{eqnarray}
$\tF_p$ is defined in eq.(\ref{surfaceG}) and $'$ means $\partial_r$. 
From the above,  
\begin{eqnarray}\label{ConsGravAnomaly}
T^r{}_{t(H)} = a_H + \int_{r_+}^r dr \partial_r \left( c_0 A_t + \frac{e^2}{4\pi}A_t^2 + N_t^r \right), 
\end{eqnarray}
where $a_H$ stands for the value of $T^r{}_{t}$ at $r=r_+$.

Now we fix $a_H$. To this purpose, we consider eq.(\ref{Anomaly2}) given by not the consistent energy-momentum tensor but the covariant energy-momentum tensor~\cite{Bardeen:1984pm}. 
Namely, we consider 
\begin{eqnarray}\label{CGA eq in H1}
\nabla_\mu \tT^\mu{}_{\nu(H)} = 
F_{\mu\nu}J^\mu_{(H)} 
+ A_\nu \nabla_\mu J^\mu_{(H)} 
\mp \frac{1}{96\pi \sqrt{-g}}\epsilon_{\mu\nu}\partial^\mu R. 
\end{eqnarray}  
Then in the region ${\cal H}$, for the chiral situation that there are only the left-handed modes, 
the identity of eq.(\ref{CGA eq in H1}) leads to
\begin{eqnarray}\label{CovaAnomaly1}
\nabla_\mu \tT^\mu{}_{\nu(H)} = F_{\mu\nu}J^\mu_{(H)} + A_\nu \nabla_\mu J^\mu_{(H)} - \frac{1}{96\pi \sqrt{-g}}\epsilon_{\mu\nu}\partial^\mu R.
\end{eqnarray} 
For the case that $\nu=t$, the above can be written as
\begin{eqnarray}\label{CovaAnomaly2}
\partial_r \tT^r{}_{t(H)} = F_{rt}J^r_{(H)} + A_t \partial_r J^r_{(H)} + \partial_r  \tN^r{}_t,
\end{eqnarray}
where $\tilde{N}^r{}_t = \frac{1}{96\pi}\big(\tF_{p} \tF_{p}''-(\tF_{p}')^2/2\big)$. 
Then we can find a relation\footnote{
In the two-dimensional gravitational anomaly, if it depends only on $r$, we can find a general relation:~$
\partial_r(\sqrt{-g}(\tilde{T}^r{}_{t(H)}-T^r{}_{t(H)}))=-\frac{1}{96\pi}
\left( \partial_r(g^{rr}R)-\partial_r g^{rr}\cdot R + \partial_r^2 \Gamma^r_{tt}) \right)$.
}:
\begin{eqnarray}\label{ConsisCovaAnomaly}
\tilde{T}^r{}_{t(H)} - T^r{}_{t(H)} = \frac{1}{192\pi}\big(\tF_p \tF_p''-2(\tF_p')^2\big). 
\end{eqnarray} 
Imposing the boundary condition $\tilde{T}^\mu{}_\nu\big|_{r=r_+}=0$, from eq.(\ref{ConsisCovaAnomaly}) we can obtain the following result:
\begin{eqnarray}\label{a_H}
a_H = \frac{(\tF_p')^2}{96\pi}\bigg|_{r=r_+} \! =\, \frac{\pi}{6\beta^2},
\end{eqnarray} 
where we have used ~$T^r{}_{t(H)}\big|_{r=r_+}=a_H$, which can be seen from eq.(\ref{ConsGravAnomaly}),  
$\beta$ means the inverse temperature given in eq.(\ref{surfaceG}).

Now considering the effective action $W$ given by the two-dimensional model, 
let us consider its variation under the general coordinate transformation, which can be written as
\begin{eqnarray}\label{dWg}
-\delta W \!\! &=& \!\! \int dtdr \sqrt{-g}\,\xi^\nu \nabla_\mu T^\mu{}_\nu \nonumber \\
          \!\! &=& \!\! \int dtdr \, \left( \xi^t \nabla_\mu T^\mu{}_t + \xi^r \nabla_\mu T^\mu{}_r \right) \nonumber \\
          \!\! &=& \!\! \int dtdr \, \xi^t \left\{ 
c_o \partial_r A_t 
+ \partial_r \bigg(\bigg( \frac{e^2}{4\pi} A_t^2 + N^r{}_t \bigg)H\bigg) 
\right. \nonumber \\ 
&& \qquad \qquad \quad \! \left.
+ \bigg( T^r{}_{t(o)} - T^r{}_{t(H)} + \frac{e^2}{4\pi}A_t^2 + N^r{}_t \bigg) \delta \big( r - (r_+ + \epsilon)\big)\right\}, 
\end{eqnarray} 
where the above is the expression obtained without the contribution of the right-handed modes as well as eq.(\ref{dWU}).   
In the r.h.s. of the third line, 
although the first term is breaking the invariance, 
it can be tolerated for the fact that our background gauge field is classical, as in eq.(\ref{2DRA})~\cite{Iso:2006wa,Iso:2006ut}. 
The second and third terms are anomalous, which should vanish, since the original theory is anomaly-free.

\textcolor{black}{The second term should be canceled by the quantum contribution of the right-handed modes.} 
Then from the demand that the third term should vanish for the anomaly-free, the following constraint is led:
\begin{eqnarray}
a_0 = a_H + \frac{m^2}{4\pi}\big(A_t (r_+)\big)^2 - N^r{}_t(r_+). 
\end{eqnarray}
Using $a_H$ in eq.(\ref{a_H}), we can obtain the following $a_0$: 
\begin{eqnarray}\label{a0_EM_result}
a_0 = \frac{\pi}{12\beta^2} + \frac{m^2}{4\pi} \left( \frac{aG^q}{r_+^2 + a^2}\right)^2 
    = \frac{(\tF_p')^2}{192\pi}\bigg|_{r=r_+} + \frac{m^2}{4\pi} \left( \frac{aG^q}{r_+^2 + a^2}\right)^2,  
\end{eqnarray} 
where $p$ and $q$ for the type II and III metrics are given in eq.(\ref{pq}). 
Since $q=0$ in the type III metric, $a_0$ in the type III metric has no dependence on the time-delay. 
Here let us notice that $m$ appearing in the above expression is not mass but an index in the partial wave expansion in the above of eq.(\ref{2DRA}). 
\newline

From the same discussion in the Appendix.A of Ref.\cite{Iso:2006ut}, 
we can see that the above $a_0$ can be identified with the total flux of energy-momentum tensor from the regular rotating black hole with a time-delay. 
To show this identification we repeat the virtually same description in Ref.\cite{Iso:2006ut} in what follows.

In general the total fluxes of the energy-momentum tensor can be computed from the rotating black holes as
\begin{eqnarray}\label{47}
m\int^\infty_0\frac{d\omega}{2\pi}\left( N_m(\omega) + N_{-m}(\omega)\right) = \frac{\pi}{12\beta^2} + \frac{m^2}{4\pi}\Omega^2, 
\end{eqnarray}
where the distribution function of fermions $N_m(\omega)$ in the rotating black holes with the time-delay is given in eq.(\ref{DistFuncFer}).
We can see that eq.(\ref{47}) is consistent with our result in eq.(\ref{a0_EM_result}).

As well as eq.(\ref{c0_AM_result}), our result involves the three effects:~
\begin{itemize}
\item $G$~(The time-delay)
\item $a$~(The angular momentum of the black hole) 
\item $\l_p$~(The quantum gravity effect regularizing our black hole singularity via $r_+(\l_p)$ as in eq.(\ref{rpm}))
\end{itemize}

We can see that when there is no time-delay~($G=1$) and no quantum gravity effect~($\l_p=0$), 
our result in eq.(\ref{a0_EM_result}) can agree with the total flux of energy-momentum tensor from the classical Kerr black hole given in eq.(36) in Ref.\cite{Iso:2006ut}. 
In addition if $a=0$ and $G=1$ with leaving $\l_p$, our result can agree with the Hawking thermal flux of Schwartzshild regular black hole given in eq.(4.7) in Ref.\cite{Kim:2008hm}.

The $r$-, $\beta$- and $m_0$-dependences in $a_0$ via $G$ can be known from Fig.\ref{Fig_G}. 
In what follows, the same comment with the last paragraph in sec.\ref{subsec:The angular momentum} will be followed. 
We skip to describe it here again.

\section{Summary}
\label{Summary} 

In this paper the Hawking fluxes  
from the rotating regular black holes with a time-delay described by the type II and III metrics has been computed based on the anomaly cancellation method.  

As for the type I metric, it has been turned out that field theories cannot be reduced to infinite two-dimensional free massless theories  
in which the anomaly cancellation method is feasible at the near-horizon region.  
Since occurring of the dimensional reduction would be natural as we have mentioned under eq.(\ref{MetricsType}),  
it could be considered that the type I metric would be some problematic metric. 
The cause for the feasibility of the dimensional reduction in the type II and III metrics and the infeasibility in the type I metric has been mentioned in Sec.\ref{Chap:DRA}.

The Hawking fluxes we have computed in this paper involve three effects as in eqs.(\ref{c0_AM_result}) and (\ref{a0_EM_result}):
\begin{itemize}
\item $G$~(The time-delay)
\item $a$~(The angular momentum of the black hole) 
\item $\l_p$~(The quantum gravity effect regularizing our black hole singularity via $r_+(\l_p)$ as in eq.(\ref{rpm}))
\end{itemize} 
\noindent
By the effects of $G$ and $\l_p$, our result could be considered to correspond to a more realistic Hawking fluxes.

%
%

What the Hawking fluxes can be obtained based on the anomaly cancellation method would be interesting 
in terms of the relation between a consistency of quantum field theories and black hole thermodynamics.
Leaving the Hawking fluxes, 
the anomaly cancellation method itself could shed new lights in resolving various problems and elucidating new aspects in the black hole thermodynamics.

Further, from the computation of the Hawking fluxes based on the anomaly cancellation method, 
we can see the fact that apart from the black holes, 
quantum tunneling processes could link to quantum anomalies. 
Of course in the actual analysis there would be several problems as whether the system is the two dimensional or not, 
and the model can reduce to an infinite free massless model like this study.
However, what there is some relation between quantum tunneling processes and quantum anomalies would be interesting.  
We might be able to get some new insight by reconsidering some quantum tunneling processes in terms of quantum anomalies.


\section*{Acknowledgment}
The author thanks 
Hiroshi Umetsu, 
Lunchakorn Tannukij, 
Rakpong Saikaew, 
Simone Speziale, 
Sujiphat Janaun 
and 
Tommaso De Lorenzo   
for helpful advise and discussion. The author also thank the Institute for Fundamental Study and Naresuan University.

\appendix

\section{Introduction to the regular black holes and its problems}
\label{app:Regularization}
Let us first write a four-dimensional non-rotating and no charged spherical black hole metric  as
\begin{eqnarray}\label{fRegBH}
ds^2 = g_{tt}dt^2 +g_{rr} dr^2 + r^2 d\Omega^2,  
\end{eqnarray} 
where $g_{rr} = - g^{-1}_{tt}$. If it comes to the Schwarzschild black hole, $- g_{tt} = 1-2m_0/r$, where $m_0$ is the ADM mass. 
However, if we attempt to get the Schwarzschild regular black hole, 
we require the metric to behave as the de Sitter-like in the central region around $0 \leq r \ll l_p$~($l_p$ is some Plank length order number with the dimension of the length)  
without changing the asymptotic behavior from the Schwarzschild black hole.    
To put it concretely, we require the following behaviors:
\begin{eqnarray}\label{AsyRegBH}
-g_{tt} = 
\left\{
\begin{array}{ll}
\displaystyle 1 - \frac{2m_0}{r}  & {\rm for} \quad r  \to \infty \\
\displaystyle 1 - \frac{r^2}{l_p^2} & \textrm{for $0 \leq r \ll l_p$}.
\end{array} 
\right.
\end{eqnarray}

Here let us give an explanation about the above. 
The quantum effect of the gravity becomes stronger in the gravity sources of the black holes when these close up to the Plank length order number each other, 
namely $r_{source} \to l_p$, where $r_{source}$ stands for the extent of the distribution of the gravity sources.    
The effect of the quantum gravity would work as the repulsive force.  
Hence, it is considered that the gravity sources are prevented from further closing up shorter than the Plank order length, and the singularities eventually do not appear.  

One way to realize such a situation is to consider that 
the space-time becomes de Sitter-like in the region around $0 \leq r \ll l_p$  
by considering the repulsive forces as the tendency of the de Sitter spaces to extend.   
Hence the situation is that the quantum gravity is working as some positive cosmological constant.
Actually, the second line in eq.(\ref{AsyRegBH}) is a solution of the Einstein equation with a positive cosmological constant:~$\Lambda=3/l_p^2$, 
and is the metric of the de Sitter space in the so-called static coordinate, 
in which there is no singularity in eq.(\ref{AsyRegBH}) for $r \to 0$ without changing the asymptotic behavior from the Schwarzschild black hole.

Then as a simple one satisfying eq.(\ref{AsyRegBH}), the following $g_{tt}$ has been proposed~\cite{Hayward:2005gi},   
\begin{eqnarray}\label{fRegBHgtt1}
-g_{tt} = 1 - \frac{2m_0 r^2}{r^3+2l_p^2m_0}.   
\end{eqnarray} 
For the above $g_{tt}$, we can see that the killing horizons can appear when $m_0 \geq m_c = 3 \sqrt{3}l_p/4$. 
Hence the Schwarzschild regular black holes could be considered by the above metric with $m_0 \geq m_c$. 

Now we change the notation in eq.(\ref{fRegBHgtt1}) to the one in Ref.\cite{Bambi:2013ufa} as
\begin{eqnarray}\label{fRegBHgtt2}
-g_{tt} = 1 - \frac{2m_0 r^2}{r^3+l_p^3}.   
\end{eqnarray} 
In this changing, only $m_0$ in the denominator of eq.(\ref{fRegBHgtt1}) is changed to $m_0 \rightarrow l_p/2$ and $m_c$ is also changed to $m_c \rightarrow 3l_p/(2\cdot 2^{2/3})$. 
But the properties to be required as the metric in the regular black holes are not changed.

Applying the Newman-Janis algorithm~\cite{Newman:1965tw,Newman:1965my,Drake:1998gf} to this regular Schwarzschild black hole metric, 
a rotating regular black hole metric has been obtained~\cite{Bambi:2013ufa}. 
However, at that time some problems appear, which we state in what follows. 
The rotating regular black hole metric we consider in this paper is the one in which such problems are cured. 
\newline

First, the asymptotic behavior of the Newton potential in the Schwarzschild regular black hole space-time with the $g_{tt}$ in eq.(\ref{fRegBHgtt2}) can be written as
\begin{eqnarray}\label{fRegP}
\Phi = -  \frac{m_0}{r} + \frac{l_p^3 m_0}{r^4} + {\cal O}(r^{-5}),  
\end{eqnarray} 
where $\Phi = - \frac{1}{2} \left(1 + g_{tt} \right)$ with the $g_{tt}$ given in eq.(\ref{fRegBHgtt2}). 
On the other hand, in the Schwarzshild and Kerr black hole metrics without the regularization, 
the Newton potentials around the asymptotic region with the 1-loop correction of the quantum gravity are given as~\cite{BjerrumBohr:2002ks} 
\begin{eqnarray}\label{fRegNewP}
\Phi_{\textrm{1-loop}} = - \frac{m_0}{r} + \frac{\gamma\, l_p^2 m_0}{ r^3}  + {\cal O}(r^{-4}),   
\end{eqnarray} 
where $\gamma$ is some factor of order $1$ associated with the 1-loop correction of the quantum gravity.  
Hence we can find that $r^{-3}$ term lacks in the Newton potential of the regular black hole in eq.(\ref{fRegBHgtt2}).

Here we can find that there is no classical contribution given by $r^{-2}$ term in both eqs.(\ref{fRegP}) and (\ref{fRegNewP}).  
As for this, we should note that Ref.\cite{BjerrumBohr:2002ks} is taking the harmonic gauge. 
If the harmonic gauge is not taken, the $r^{-2}$ term does not appear. 
Hence indeed the result written in eq.(\ref{fRegNewP}) is the one in Ref.\cite{BjerrumBohr:2002ks} without the harmonic gauge.  
So this is why the numerical factor in eq.(\ref{fRegNewP}) is not written specifically no more than being represented as $\gamma$. 
Because in Ref.\cite{BjerrumBohr:2002ks}, it is given as $62/15\pi^2$. 
But it is a result computed in the harmonic gauge, and the results we are considering now is the ones without the harmonic gauge. 
Hence since it is expected that the coefficient is changed now, we represent it just as $\gamma$.

The lacking of $r^{-3}$ term is a problem in the Schwarzschild regular black hole metric in eq.(\ref{fRegBHgtt2}), 
and there is the same problem in the rotating regular black hole metric obtained from the Schwarzschild regular black hole via the Newman-Janis algorithm. 

In addition to the problem of the lacking of $r^{-3}$ term, 
there is another problem, which is the lack of the time-delay. 
We explain it in the next appendix.

These two problems have been firstly pointed out in Ref.\cite{DeLorenzo:2014pta}. 
In the next appendix, we show the Schwarzschild regular black hole metric in which these two problems are cured.

\section{Rotating regular black hole with a time-delay and the correct 1-loop quantum correction}
\label{app:TimeDelay}
We have pointed out the two problems in the preceding appendix, which are the lacks of the $r^{-3}$ term and the time-delay in the Schwarzschild regular black hole metric with $g_{tt}$ in eq.(\ref{fRegBHgtt2}).   
In this appendix we give the rotating regular black hole in which these two problems are cured according to Ref.\cite{DeLorenzo:2015taa}.   
\newline

Regarding what's the time-delay, a clock in the center of a cloud of dust shows a time shorter than that of a clock put at infinity for the gravitational effect of the black hole geometry. 
This is the time-delay, which should always present and is indispensable if it is a black hole geometry~\cite{Shapiro:1964uw}.     
To put it concretely, writing the times counted in the clocks put in the asymptotic region and the center of the regular black holes as $\delta t_\infty$ and $\delta t_0$ respectively,  
the time-delay can be written as
\begin{eqnarray}\label{TimeDifference} 
\frac{\delta t_\infty - \delta t_0}{\delta t_\infty} = 1-\sqrt{\big|g_{tt}(r=0)\big|} \,\,\in\,\, [0,~1),  
\end{eqnarray} 
where the time-delay gets smaller/larger, when r.h.s. closes to zero/one, respectively.

Then we can find that there is no time-delay in the regular black holes with $g_{tt}$ in eq.(\ref{fRegBHgtt2}). 
This is because the geometries at the center of the regular black holes are flat as can be seen from eq.(\ref{AsyRegBH}), 
and this is the problem of the lack of the time-delay in the regular black holes.    
In order to improve this situation, a replacement:~$g_{tt} \to G g_{tt}$ at the stage of the Schwarzschild regular black holes has firstly been proposed~\cite{DeLorenzo:2014pta}, 
where $g_{tt}$ is the one in eq.(\ref{fRegBHgtt2}) and $G = 1 - \frac{\beta m_0 \alpha}{\alpha r^3 + \beta m_0}$. 
Here we should note that this $G$ is not the $G$ we finally use in this paper.

By this, we can retrieve the time-delay in the regular Schwrtzshild black hole.
Furthermore, by this we can find that the another problem, the lack of the $r^{-3}$ term, could also be cured. 
Hence now it looks that the two problems can be cured.  
However it turns out that the curvature diverges 
in the rotating regular black hole metric obtained from this improved Schwarzschild regular black hole metric 
via the Newman-Janis algorithm~\cite{Newman:1965tw,Newman:1965my,Drake:1998gf}.  
In such a situation, in Ref.\cite{DeLorenzo:2015taa}, 
another version of $G$ given in eq.(\ref{GTD}) has been newly proposed with the procedure above eq.(\ref{GTD}).  
\newline

We now show the behavior of $Gg_{tt}$ obtained from the procedure above eq.(\ref{GTD}) in the neighborhood of $r = 0$ as
\begin{eqnarray}\label{Ggttalpha} 
Gg_{tt} &=& \left(1 - \alpha + {\cal O}\left(r^4\right)\right) + \left(\alpha + {\cal O}\left(r^4\right)\right) e^{-\frac{\beta  m_0}{\alpha  r^3}} \nonumber\\ 
        &=& 1-\alpha + {\cal O}(r^4),
\end{eqnarray}  
where now $G$ and $g_{tt}$ in the above are the ones in eqs.(\ref{GTD}) and (\ref{RRBHTD}), respectively.
Considering the above $Gg_{tt}$ as a $g_{tt}(r=0)$ in eq.(\ref{TimeDifference}), 
we can see that the time-delay could be involved in the rotating regular black hole without the divergence of the curvature~\cite{DeLorenzo:2015taa}.   
Here the time-delay gets larger/smaller for smaller/larger $\alpha\, \in\, [0,~1)$, respectively.

Further we can see that the $r^{-3}$ term suggested in the 1-loop quantum gravity~\cite{BjerrumBohr:2002ks}  
can appear in the Newton potential computed from this $Gg_{tt}$ as  
\begin{eqnarray}
\Phi = - \frac{m_0}{r} - \frac{m_0 \left( \beta - 2 a^2 \cos^2\theta \right)}{2r^3} + \frac{m_0 \left( l_p^3 + m_0 \beta \right)}{r^4} +{\cal O}(r^{-5}). 
\end{eqnarray}

Hence now we can consider that the rotating regular black hole with a time-delay 
in which the two problems~(the lacks of the $r^{-3}$ term in the 1-loop Newton potential and the time-delay without the divergence of the curvature) are cured 
can be safely obtained by the procedures written above above eq.(\ref{GTD}).

\section{Derivations of eqs.(\ref{dWU}) and (\ref{dWg})}
\label{app:DimRed}

In this appendix, we roughly show how the first lines in eqs.(\ref{dWU}) and (\ref{dWg}) can be derived. 
 
\subsection{Derivation of eq.(\ref{dWU})}

First taking some fields $\phi_i(x)$, let's consider its global transformation,
\begin{eqnarray} \label{GlobalTrans}
\phi_i(x) \rightarrow \phi_i'(x)=\phi_i(x)+\epsilon \, G_i(\phi_i(x))
\end{eqnarray} 
where $\epsilon$ is some infinitesimal constant. 
Then we consider the variation of the action under this transformation as
\begin{eqnarray} 
\delta S 
&=& 
\int d^4x \left( 
\frac{\partial{\cal L}}{\partial \phi_i} \epsilon \, G_i 
+ 
\frac{\partial{\cal L}}{\partial (\partial_\mu \phi_i)} \partial_\mu \left( \epsilon \, G_i \right) 
\right)\nonumber\\
&=& 
\epsilon \int d^4x \left( 
\frac{\partial{\cal L}}{\partial \phi_i} G_i 
+ 
\frac{\partial{\cal L}}{\partial( \partial_\mu \phi_i)} \partial_\mu G_i 
\right).
\end{eqnarray} 
At this time if the integrand above can be written as a total derivative as 

\begin{eqnarray}
\frac{\partial{\cal L}}{\partial \phi_i} G_i 
+ 
\frac{\partial{\cal L}}{\partial( \partial_\mu \phi_i)} \partial_\mu G_i=\partial_\mu X^\mu(\phi), 
\end{eqnarray} 
the action can become invariant under the transformation in eq.(\ref{GlobalTrans}).

Then under the situation that such a global transformation can hold with some function $X^\mu(\phi)$, 
let us consider a local transformation depending on a infinitesimal transformation parameter $\epsilon(x)$. 
Then the variation of the action can be written as  
\begin{eqnarray}
\delta S
&=&
\int d^4x \left\{
\frac{\partial{\cal L}}{\partial \phi_i}\epsilon(x) G_i
+
\frac{\partial{\cal L}}{\partial (\partial_\mu \phi_i)} \partial_\mu \left( \epsilon(x) G_i \right) 
\right\}\nonumber\\
&=&
\int d^4x \,  \partial_\mu\epsilon(x) 
\left\{
\frac{\partial{\cal L}}{\partial \left( \partial_\mu \phi_i \right)} G_i
-X^\mu(\phi)
\right\}\nonumber\\
&=&
-\int d^4x \, \epsilon(x) \, \partial_\mu
\left\{
\frac{\partial{\cal L}}{\partial \left( \partial_\mu \phi_i \right)} G_i
-X^\mu(\phi)
\right\}.
\end{eqnarray} 
$\{ \cdots \}$ is the conserved current, and we can see that this is the same form with the first line in eq.(\ref{dWU}).  
 
\subsection{Derivation of eq.(\ref{dWg})} 

Now let us consider an action written as $S(A_i(x^m),g_{ij}(x^m))$, 
where $A_i(x^m)$ and $g_{ij}(x^m)$ mean some matter fields and metrics. 
Then let us consider the general coordinate transformation, 
\begin{eqnarray}\label{DiiffeoCordTrans} 
x^m \rightarrow x'{}^m=x^m+ \xi^m(x).
\end{eqnarray} 
We write the transformation of fields under this general coordinate transformation as 
\begin{eqnarray}
A_i(x^m) \rightarrow A'_i(x^m) &=& A_i(x^m) + \delta A_i(x^m),\\
g_{ij}(x^m) \rightarrow g'_{ij}(x^m) &=& g_{ij}(x^m) + \delta g_{ij}(x^m).
\end{eqnarray} 
Then under the transformation in eq.(\ref{DiiffeoCordTrans}), the variation of the action can be written as
\begin{eqnarray} 
\delta S &=& S(A_i+\delta A_i,g_{ij}+\delta g_{ij})- S(A_i,g_{ij})\nonumber\\
         &=& \delta S\big|_{\textrm{$g_{ij}$ fixed}} + \delta S\big|_{\textrm{$A_i$ fixed}}, \label{deltaS}
\end{eqnarray}
where
\begin{eqnarray}
\delta S\big|_{\textrm{$g_{ij}$ fixed}} &=& S(A_i+\delta A_i,g_{ij} + \delta g_{ij}) - S(A_i,g_{ij}+ \delta g_{ij}), \\
\delta S\big|_{\textrm{$A_i$ fixed}} &=& S(A_i,g_{ij}+ \delta g_{ij}) - S(A_i,g_{ij}).
\end{eqnarray}
The theory should be invariant under the general coordinate transformation in eq.(\ref{DiiffeoCordTrans}). 
So we demand $\delta S=0$. 
Here $\delta S\big|_{\textrm{$g_{ij}$ fixed}}$ should vanish for a normal assumption that $A_i$ satisfies the equations of motion. 
Therefore eq.(\ref{deltaS}) becomes
\begin{eqnarray}\label{DiiffeoCordTrans2}
\delta S = \delta S\big|_{\textrm{$A_i$ fixed}}\,.
\end{eqnarray}

$\delta S\big|_{\textrm{$A_i$ fixed}}$ can be generally written as 
\begin{eqnarray}\label{DiiffeoCordTrans2}
\delta S\big|_{\textrm{$A_i$ fixed}} 
&=& \int d^4x 
\left\{ 
\frac{\partial(\sqrt{-g}{\cal L})}{\partial g^{ij}} \delta g^{ij} 
+  
\frac{\partial(\sqrt{-g}{\cal L})}{\partial\left( \frac{\partial g^{ij}}{\partial x^k} \right)} \delta \left( \frac{\partial g^{ij}}{\partial x^k}\right) 
\right\}\nonumber\\
&=& \int d^4x 
\left\{ 
\frac{\partial(\sqrt{-g}{\cal L})}{\partial g^{ij}} \delta g^{ij} 
- 
\frac{\partial}{\partial x^k}
\left(
\frac{\partial(\sqrt{-g}{\cal L})}{\partial\left( \frac{\partial g^{ij}}{\partial x^k} \right)} 
\right) \delta g^{ij}
\right\}\nonumber\\
&=& -\frac{1}{2} \int d^4x \sqrt{-g} \, T_{ij} \, \delta g^{ij},
\end{eqnarray}
where from the first line to the second line a partial derivative has been performed with $\delta \left( {\partial g^{ij}}/{\partial x^k}\right) = {\partial \,\delta g^{ij}}/{\partial x^k} $, 
and in the last line 
$
T_{ij}
=
\frac{2}{\sqrt{-g}}\left\{ 
\frac{\partial}{\partial x^k} \left(\frac{\partial(\sqrt{-g}{\cal L})}{\partial\left( \frac{\partial g^{ij}}{\partial x^k} \right)} \right)
-\frac{\partial(\sqrt{-g}{\cal L})}{\partial g^{ij}} \right\}
$ means the energy-momentum tensor. 

Now $\delta g^{ij}=\nabla^j \xi^i + \nabla^i \xi^j$. Then
\begin{eqnarray}
\textrm{Eq.(\ref{DiiffeoCordTrans2})} &=& -\frac{1}{2} \int d^4x \sqrt{-g} \, T_{ij} \left( \nabla^j \xi^i + \nabla^i \xi^j \right) \nonumber\\
                                      &=& -\int d^4x \sqrt{-g} \, T_{ij} \nabla^j \xi^i \nonumber\\
                                      &=& -\int d^4x \sqrt{-g} \, \left\{ \nabla_j(T^{ij} \xi_i) - \xi_i \nabla_j T^{ij}  \right\}.
\end{eqnarray}
Since the first term  above can be rewritten into $\nabla_j(T^{ij} \xi_i)=\frac{1}{\sqrt{-g}} \frac{\partial}{\partial x^j}(\sqrt{-g}T^{ij} \xi_i)$, 
the contribution from the first term becomes zero. 
Therefore finally
\begin{eqnarray}
\textrm{Eq.(\ref{DiiffeoCordTrans2})} &=& \int d^4x \sqrt{-g} \, \xi_i  \nabla_j T^{ij} .
\end{eqnarray}
This is the same form with the first line in eq.(\ref{dWg}).


\begin{thebibliography}{100} 




\bibitem{DeLorenzo:2015taa} 
  T.~De Lorenzo, A.~Giusti and S.~Speziale,
  ``Non-singular rotating black hole with a time delay in the center,''
  Gen.\ Rel.\ Grav.\  {\bf 48}, no. 3, 31 (2016)
  [arXiv:1510.08828 [gr-qc]].

\bibitem{Bardeen1968} 
J. M. Bardeen. in Conference Proceedings of GR5 (Tbilisi, USSR, 1968), p. 174.

\bibitem{Dymnikova:1992ux} 
  I.~Dymnikova,
  ``Vacuum nonsingular black hole,''
  Gen.\ Rel.\ Grav.\  {\bf 24}, 235 (1992).

\bibitem{AyonBeato:1998ub} 
  E.~Ayon-Beato and A.~Garcia,
  ``Regular black hole in general relativity coupled to nonlinear electrodynamics,''
  Phys.\ Rev.\ Lett.\  {\bf 80}, 5056 (1998)
  [gr-qc/9911046].

\bibitem{Cho:2000kr} 
  H.~Cho, D.~Kastor and J.~H.~Traschen,
  ``The Dynamics of collapsing monopoles and regular black holes,''
  hep-th/0002220.
  
\bibitem{Dymnikova:2001fb} 
  I.~Dymnikova,
  ``Cosmological term as a source of mass,''
  Class.\ Quant.\ Grav.\  {\bf 19}, 725 (2002)
  [gr-qc/0112052]. 

\bibitem{Hayward:2005gi} 
  S.~A.~Hayward,
  ``Formation and evaporation of regular black holes,''
  Phys.\ Rev.\ Lett.\  {\bf 96}, 031103 (2006)
  [gr-qc/0506126].

\bibitem{Nicolini:2005vd} 
  P.~Nicolini, A.~Smailagic and E.~Spallucci,
  ``Noncommutative geometry inspired Schwarzschild black hole,''
  Phys.\ Lett.\ B {\bf 632}, 547 (2006)
  [gr-qc/0510112].

\bibitem{Ansoldi:2006vg} 
  S.~Ansoldi, P.~Nicolini, A.~Smailagic and E.~Spallucci,
  ``Noncommutative geometry inspired charged black holes,'' 
  Phys.\ Lett.\ B {\bf 645}, 261 (2007)
  [gr-qc/0612035].

\bibitem{DeLorenzo:2014pta} 
  T.~De Lorenzo, C.~Pacilio, C.~Rovelli and S.~Speziale,
  ``On the Effective Metric of a Planck Star,''
  Gen.\ Rel.\ Grav.\  {\bf 47}, no. 4, 41 (2015)
  [arXiv:1412.6015 [gr-qc]].

\bibitem{Bambi:2013ufa} 
  C.~Bambi and L.~Modesto,
  ``Rotating regular black holes,''
  Phys.\ Lett.\ B {\bf 721}, 329 (2013)
  [arXiv:1302.6075 [gr-qc]]. 

\bibitem{Toshmatov:2014nya} 
  B.~Toshmatov, B.~Ahmedov, A.~Abdujabbarov and Z.~Stuchlik, 
  ``Rotating Regular Black Hole Solution,''
  Phys.\ Rev.\ D {\bf 89}, no. 10, 104017 (2014)
  [arXiv:1404.6443 [gr-qc]].

\bibitem{Ghosh:2014hea} 
  S.~G.~Ghosh and S.~D.~Maharaj,
  ``Radiating Kerr-like regular black hole,''
  Eur.\ Phys.\ J.\ C {\bf 75}, 7 (2015) 
  [arXiv:1410.4043 [gr-qc]].


\bibitem{Neves:2014aba} 
  J.~C.~S.~Neves and A.~Saa,
  ``Regular rotating black holes and the weak energy condition,''
  Phys.\ Lett.\ B {\bf 734}, 44 (2014) 
  [arXiv:1402.2694 [gr-qc]].  

%
%

\bibitem{Shapiro:1964uw} 
  I.~I.~Shapiro,
  ``Fourth Test of General Relativity,''
  Phys.\ Rev.\ Lett.\  {\bf 13}, 789 (1964).
  

\bibitem{Umetsu:2010ts} 
  K.~Umetsu, 
  ``Recent Attempts in the Analysis of Black Hole Radiation,''
  arXiv:1003.5534 [hep-th].
  

\bibitem{Robinson:2005pd} 
  S.~P.~Robinson and F.~Wilczek,
  ``A Relationship between Hawking radiation and gravitational anomalies,''
  Phys.\ Rev.\ Lett.\  {\bf 95}, 011303 (2005) 
  [gr-qc/0502074].
  
\bibitem{Iso:2006wa}  
  S.~Iso, H.~Umetsu and F.~Wilczek,
  ``Hawking radiation from charged black holes via gauge and gravitational anomalies,''
  Phys.\ Rev.\ Lett.\  {\bf 96}, 151302 (2006)
  [hep-th/0602146].
 
\bibitem{Iso:2006ut}  
  S.~Iso, H.~Umetsu and F.~Wilczek,
  ``Anomalies, Hawking radiations and regularity in rotating black holes,''
  Phys.\ Rev.\ D {\bf 74}, 044017 (2006)   
  [hep-th/0606018].

\bibitem{Banerjee:2007uc}  
  R.~Banerjee and S.~Kulkarni,
  ``Hawking radiation, effective actions and covariant boundary conditions,''
  Phys.\ Lett.\ B {\bf 659}, 827 (2008)
  [arXiv:0709.3916 [hep-th]].
  
  
\bibitem{Banerjee:2007qs}  
  R.~Banerjee and S.~Kulkarni,
  ``Hawking radiation and covariant anomalies,''
  Phys.\ Rev.\ D {\bf 77}, 024018 (2008)
  [arXiv:0707.2449 [hep-th]].
  
  
\bibitem{Bardeen:1984pm} 
  W.~A.~Bardeen and B.~Zumino,
  ``Consistent and Covariant Anomalies in Gauge and Gravitational Theories,''
  Nucl.\ Phys.\ B {\bf 244}, 421 (1984).


\bibitem{Kim:2008hm} 
  W.~Kim, H.~Shin and M.~Yoon,
  ``Anomaly and Hawking radiation from regular black holes,'' 
  arXiv:0803.3849 [gr-qc].
    
\bibitem{Newman:1965tw} 
  E.~T.~Newman and A.~I.~Janis,
  ``Note on the Kerr spinning particle metric,''
  J.\ Math.\ Phys.\  {\bf 6}, 915 (1965).

\bibitem{Newman:1965my} 
  E.~T.~Newman, R.~Couch, K.~Chinnapared, A.~Exton, A.~Prakash and R.~Torrence,
  ``Metric of a Rotating, Charged Mass,''
  J.\ Math.\ Phys.\  {\bf 6}, 918 (1965).

\bibitem{Drake:1998gf} 
  S.~P.~Drake and P.~Szekeres,
  ``Uniqueness of the Newman-Janis algorithm in generating the Kerr-Newman metric,''
  Gen.\ Rel.\ Grav.\  {\bf 32}, 445 (2000)
  [gr-qc/9807001].

\bibitem{Bertlmann:Anomalies}  R.~Bertlmann, Anomalies In Quantum Field Theory (Oxford Science Publications, Oxford,

\bibitem{Fujikawa:2004}  K.~Fujikawa and H.~Suzuki, Path integrals and quantum anomalies (Oxford Science Publications, Oxford, 2004).


\bibitem{Parikh:1999mf} 
  M.~K.~Parikh and F.~Wilczek,
  ``Hawking radiation as tunneling,''
  Phys.\ Rev.\ Lett.\  {\bf 85}, 5042 (2000)
  [hep-th/9907001].  
  
\bibitem{BjerrumBohr:2002ks} 
  N.~E.~J.~Bjerrum-Bohr, J.~F.~Donoghue and B.~R.~Holstein,
  ``Quantum corrections to the Schwarzschild and Kerr metrics,''
  Phys.\ Rev.\ D {\bf 68}, 084005 (2003)
  [Phys.\ Rev.\ D {\bf 71}, 069904 (2005)]
  [hep-th/0211071].
  
  
  
  
  
  
  

%
%
%
%
   
\end{thebibliography}
\end{document}